\documentclass[journal=jpccck,manuscript=article]{achemso}

\author{V.~R.~Galakhov}
\email{galakhov@ifmlrs.uran.ru}
\affiliation[Institute of Metal Physics]{Institute of Metal Physics, Russian Academy of Sciences --- Ural
Division, S. Kovalevskaya str., 18, 620990 Yekaterinburg, Russia}
\alsoaffiliation[Ural State Mining University]{Ural State Mining University, Kuibyshev str., 30, 620144 Yekaterinburg, Russia}

\author{A.~Buling}
\affiliation[Universit\"at Osnabr\"uck]
{Fachbereich Physik, Universit\"at Osnabr\"uck, Barbarastrasse 7,
D-49069 Osnabr\"uck, Germany}

\author{M.~Neumann}
\affiliation[Universit\"at Osnabr\"uck]
{Fachbereich Physik, Universit\"at Osnabr\"uck, Barbarastrasse 7,
D-49069 Osnabr\"uck, Germany}

\author{N.~A.~Ovechkina}  
\affiliation[Institute of Metal Physics]{Institute of Metal Physics, Russian Academy of Sciences --- Ural
Division, S. Kovalevskaya str., 18, 620990 Yekaterinburg, Russia}

\author{A.~S.~Shkvarin}  
\affiliation[Institute of Metal Physics]{Institute of Metal Physics, Russian Academy of Sciences --- Ural
Division, S. Kovalevskaya str., 18, 620990 Yekaterinburg, Russia}

\author{A.~S.~Semenova}
\affiliation[Institute of Solid State Chemistry]{Institute of Solid State Chemistry, Russian Academy of Sciences
--- Ural Division, Pervomayskaya~str.,~91, 620990 Yekaterinburg, Russia}

\author{M.~A.~Uimin}  
\affiliation[Institute of Metal Physics]{Institute of Metal Physics, Russian Academy of Sciences --- Ural
Division, S. Kovalevskaya str., 18, 620990 Yekaterinburg, Russia}

\author{A.~Ye.~Yermakov}  
\affiliation[Institute of Metal Physics]{Institute of Metal Physics, Russian Academy of Sciences --- Ural
Division, S. Kovalevskaya str., 18, 620990 Yekaterinburg, Russia}

\author{E.~Z.~Kurmaev}  
\affiliation[Institute of Metal Physics]{Institute of Metal Physics, Russian Academy of Sciences --- Ural
Division, S. Kovalevskaya str., 18, 620990 Yekaterinburg, Russia}

\author{O.~Y.~Vilkov}  
\affiliation[Technische Universit\"at Dresden]{Technische Universit\"at Dresden, Germany}
\alsoaffiliation[St. Petersburg State University]{St. Petersburg State University, Russia}

\author{D.~W.~Boukhvalov}  
\affiliation[School of Computational Sciences]{School of Computational Sciences, 
Korea Institute for Advanced Study (KIAS) Hoegiro 87, Dongdaemun-Gu, Seoul, 130-722, Korean Republic
}

\title[Carbon States in Carbon-Encapsulated Nickel 
 Nanoparticles Studied by Means of X-Ray Absorption, Emission, and Photoelectron Spectroscopies]
{Carbon States in Carbon-Encapsulated Nickel 
 Nanoparticles Studied by Means of X-Ray Absorption, Emission, and Photoelectron Spectroscopies}

\begin{document}

\begin{abstract}

Electronic structure of nickel nanoparticles encapsulated in carbon was characterized by 
photoelectron, X-ray absorption, and X-ray emission spectroscopies. 
Experimental spectra are compared with the density of states calculated in the frame of the
 density functional theory. 
The carbon shell of Ni nanoparticles has been found to be multilayer graphene
with significant (about 6\%) amount of Stone--Wales defects.
Results of the experiments evidence protection of
the metallic nanoparticles from the environmental 
degradation by providing a barrier against oxidation at least for two years. 
Exposure in air for 2 years leads to oxidation only of the carbon shell of Ni@C nanoparticles
with coverage of functional groups.

\end{abstract}

\maketitle
\section{Introduction}\label{sec:intro}
Metallic nanoparticles have potential applications in many technical and biological fields 
(drug  delivery, cancer diagnostics, catalysis  etc.). 
\cite{Wu-Immunofluorescent-02,Paciott-04-nanoparticle,Alivisatos-04-nano-bio,Seo_FeCo-C-nano-2006,Hayashi-06-carbon-nanotubes,Elias-05-FeCo-nanowires,Leonhardt-03-nanotubes,Choi-03-Cu-nanowires,Bao-02-nanotubes,Bao-02-Ni-nano-magn,Scott-95-nano-morphology} 
In order to prevent metallic nanoparticles from oxidation and agglomerations upon thermal treatments,
they are encapsulated in carbon \cite{Ruoff_LaC-nature-1993,lokteva-2009,ermakov-RJPC-2009,Huang-NiC-2009}.
 Metallic nanoparticles wrapped in carbon layers can retain the intrinsic properties of metals at nanoscales.

Magnetic properties of these systems depend on sizes of the particles as well as 
their degree of crystallinity and the chemical state. 
Carbon-coated ferromagnetic metal nanoparticles are predominantly single domain and 
display the superparamagnetic behavior \cite{Scott-95-nano-morphology}. 
Moreover, the carbon coatings can endow magnetic particles with biocompatibility 
and stability in many organic and inorganic media
\cite{Hayashi-06-carbon-nanotubes,Elias-05-FeCo-nanowires,Leonhardt-03-nanotubes,Choi-03-Cu-nanowires,Bao-02-nanotubes,
ermakov-RJPC-2009}. The instability, large aggregation, and rapid biodegradation of the pure uncoated (naked) 
magnetic nanoparticles could be overcome by replacement of the naked magnetic nanoparticles 
by the encapsulated ones, since the intrinsic properties of the inner magnetic cores in this case 
are well protected by outer carbon shells, which in turn gives many possibilities for 
further modification.
Besides, the combination of carbon coating and metallic  cores having `core-shell' structure 
provides unusual catalytic properties to these nanocomposites  \cite{lokteva-2009,ermakov-RJPC-2009}. 
However, up to now, the nature of the catalytic activity of the mentioned nanoparticles is still an open question. 

Initially, we have studied carbon encapsulated nickel and iron nanoparticles by means of X-ray absorption 
and photoelectron spectroscopy methods \cite{Galakhov-JPCC-2010}. 
We have found that Fe and Ni cores are in metallic states and carbon coating protects the metallic 
nanoparticles from the environmental degradation. 
In the present paper we have concentrated the main attention for study of Ni@C nanoparticles. 
The previous studies are completed 
(i) by measurements of X-ray emission spectra (XES), 
(ii) by studies of Ni@C nanoparticles after 2 years storage, 
(iii) by calculations of the electronic structure of Ni@C using the density functional 
theory (DFT) for different models of carbon coating.
We will show that the carbon shell of Ni nanoparticles can be described by multilayer graphene
with some amount of Stone--Wales defects.

\section{Experimental and calculation details}\label{sec:experimental}
Nanocomposites Ni@C were produced by evaporation of overheated  ($\sim 2000~{^{\circ}}$C) Ni liquid drop by 
blowing up a buffer inert gas (Ar) containing a hydrocarbon as described in our earlier paper \cite{Galakhov-JPCC-2010}. 
The melting of Ni drop performed by induction levitation system which provides clean conditions  
(no impurities and oxygen) for  the obtained nanocomposites. 
The morphological characteristics and the structural parameters of the composites were analyzed by 
high-resolution transmission electron microscopy.  
Ni@C particles sizes are in the range of a few nanometers up to about 20~nm.   

Our first spectral measurements were carried out within 2 days after preparing the 
Ni@C sample  \cite{Galakhov-JPCC-2010}.  
Replicates of the experiments were made after two years on the same Ni@C sample.

The C K$\alpha$ X-ray  emission spectrum (XES) was measured at the beamline 8.0.1 of the Advanced 
Light Source (ALS) at the Lawrence Berkeley National Laboratory. 
Photons were delivered to the sample via the beamline 
undulator insertion device and the spherical grating monochromator. 

The measurements of X-ray absorption (XAS) and photoelectron (XPS) spectra 
of  Ni@C were performed using the Multi User Stage for Angular Resolved Photoemission 
(MUSTANG) experimental station at the Russian--German beamline at BESSY equipped with a 
PHOIBOS 150 SPECS GmbH electron energy analyzer. 
The  absorption spectra are registered in the total electron yield mode.
The Ni~2p  and C~1s XPS  spectra for the as-prepared Ni@C sample [denoted as Ni@C (1)] 
were obtained at the photon energy of 950~eV. 
The valence-band photoelectron spectrum of Ni@C~(1) was measured at the excitation energy of 200~eV. 

X-ray photoelectron spectra for the Ni@C sample stored in air for 2 years [Ni@C (2)]  were measured 
with an ESCA spectrometer from Physical Electronics (PHI~5600~ci) using monochromatic
Al~$K\alpha $ radiation ($h\nu = 1486.6$~eV).  
 
We have measured also spectra of reference samples: Ni metal, NiO and highly oriented pyrolytic graphite (HOPG).

The calculations of the electronic structures of Ni@C nanoparticles
were carried out with help of the density functional 
theory (DFT) using the pseudo potential SIESTA code \cite{SIESTA,Soler-31}, within the framework of the 
local density approximation (LDA) \cite{perdew_1981}. 
For the simulation of the atomic structure Ni@C nanoparticles, we have performed optimization of 
the geometric structure for several model systems, and then compared the obtained electronic structure 
with the experimental spectra. 

\section{Results and discussion}\label{sec:results}

\begin{figure}[h]
    	\includegraphics[width=0.55\textwidth,angle=270]{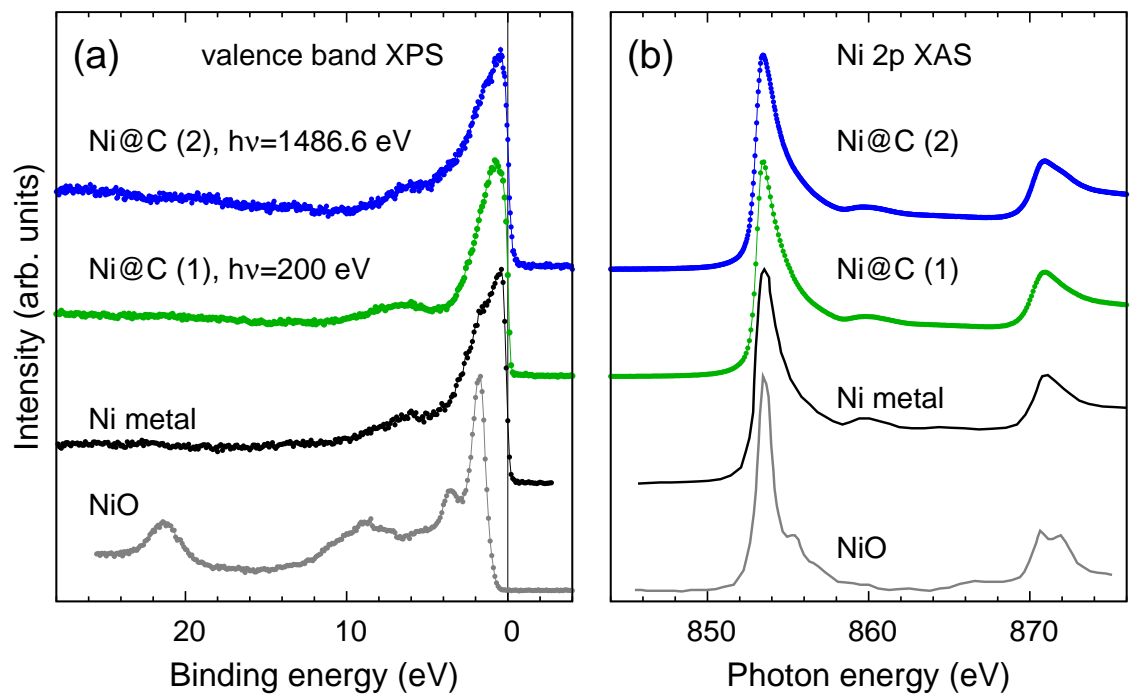}
    	\caption{(a) Valence-band photoelectron spectra of as-prepared Ni@C nanoparticles (1) 
    	and Ni@C stored in air for 2 years (2) measured at the excitation energies 200 eV and 1486.6 eV, respectively.
    	For comparison, the spectra of pure metals and oxides are shown. 
    	The spectra of Ni metal and NiO were measured at the excitations energy of 1486.6 eV. \protect\\
    	(b) Ni 2p X-ray absorption spectra of  Ni@C nanoparticles, metallic Ni, and NiO. 
    	}
    	\label{f:valence-band-Ni2p-XPS-NiC}
\end{figure}

Photoelectron spectroscopy is a feasible method for characterization of chemical states of materials and for providing 
information about the occupied electronic states.   
\ref{f:valence-band-Ni2p-XPS-NiC} (a) shows valence-band photoelectron spectra 
measured for the as-prepared Ni@C nanoparticles  [Ni@C (1)]   (with excitation energy of 200 eV) 
and for the same sample stored in air for 2 years [Ni@C (2)] 
(with excitations energy of 1486.6 eV --- Al K$\alpha$ excitation line).  
For comparison, the spectra of Ni metal and NiO oxides measured at 1486.6 eV are shown. 

According to data of Yeh and Lindau\cite{yeh85},  the electron photo-ionization cross-section ratios for Ni and C for 
the excitation energies of 200 eV and 1486.6~eV  are equal to 
110 and 590, respectively.
Consequently the main contribution to the valence-band XPS spectrum in the presented energy region 
stems from the Ni 3d states.
The similarity of the spectra of Ni metal and Ni@C demonstrates that Ni cores in Ni@C are in the metallic state.

In transition 3d metal compounds, cation 2p X-ray absorption spectra are dominated by 
intra-atomic and short-range effects. 
In view of this, metal 2p X-ray absorption spectra  correspond to the  
metal 2p$\rightarrow {\rm metal~valence~band}$~(3d) transitions and are determined by the
valence state of metal atoms. 
\ref{f:valence-band-Ni2p-XPS-NiC} (b) displays Ni 2p X-ray  absorption spectra of the 
carbon-encapsulated Ni nanoparticles, Ni metal, and NiO.
The Ni 2p X-ray  absorption spectra of the  carbon-encapsulated Ni nanoparticles are similar to 
those of metallic  Ni.
This means that Ni cores of as-prepared Ni@C (1)  and  Ni@C nanoparticles  stored in air for 2 years (2) 
are in the metallic state.

These results demonstrate high efficiency of carbon protecting layers. Note, Cho et al. \cite{Cho_Fe-Au-2004} found that 
Au-coated Fe nanoparticles are oxidized after one month.
Therefore carbon protects 3d metal against oxidation much better than gold. 

\begin{figure}[h]
    	\includegraphics[width=0.55\textwidth,angle=0]{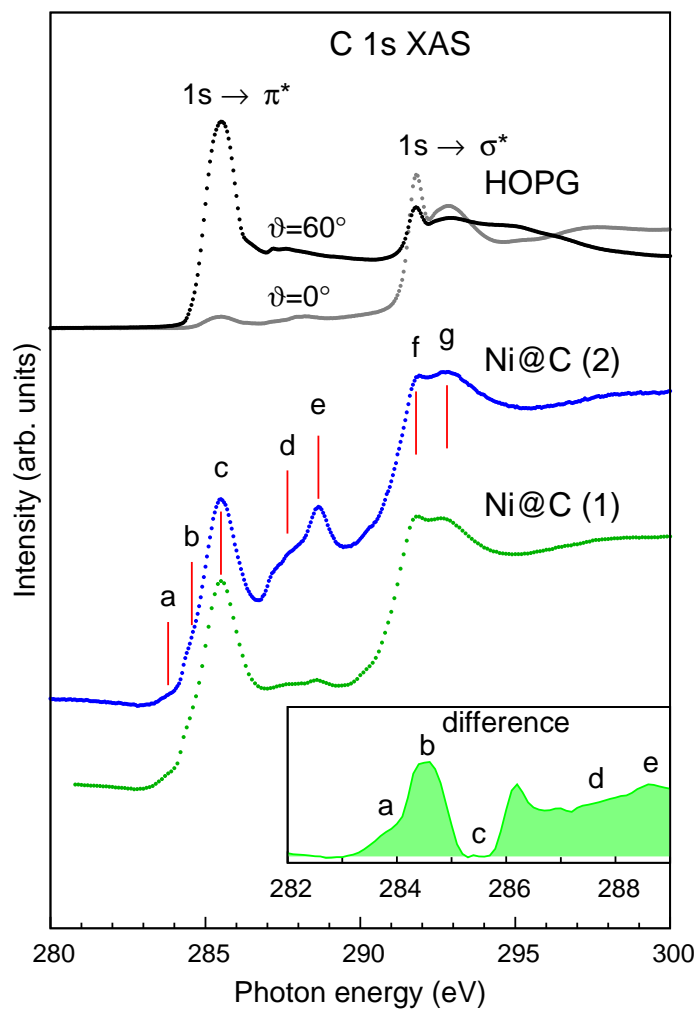}
    	\caption{C~1s X-ray absorption spectra of HOPG and of as-prepared Ni@C nanoparticles (1) 
    	and Ni@C stored in air for 2 years (2). The spectra of HOPG are
    	measured at two angles $\vartheta $ between the normal to the surface of 
    	the samples and the X-ray beam. Insert  shows the difference spectrum obtained by 
    	subtraction of the HOPG spectrum ($\vartheta = 60^{\circ}$) 
    	from the spectrum of as-prepared Ni@C. 
    	}
    	\label{f:C1s-XAS-NiC-multi}
	\end{figure}

Information about unoccupied carbon 2p states can be obtained from C~1s X-ray absorption spectra. 
C~1s X-ray absorption spectra of carbon encapsulated nanoparticles Ni@C and of HOPG are 
presented in \ref{f:C1s-XAS-NiC-multi}. 
The  spectra of HOPG are measured at two angles between the normal to the surface and of the synchrotron beam 
($\vartheta =60^{\circ}$ and $\vartheta =0^{\circ}$). 
The measurements of the C~1s XAS spectra of HOPG at two orientations allow to separated $\pi^* $ 
and $\sigma^*$ states.
The spectrum measured at $\vartheta =0^{\circ}$ reveals mainly $\sigma^*$ states.
The $\pi^*$ states are seen in the spectrum measured at $\vartheta =60^{\circ}$.

The C~1s XAS spectra of Ni@C are generally very similar to those of
graphite, showing distinct $\pi^*$ and $\sigma^*$ bands (peaks {\em c}, {\em f}, and {\em g}), 
but with some noticeable 
exceptions.
The spectra of Ni@C show peaks {\em a}, {\em b}, {\em d}, and {\em e}  which are not exhibited in 
the spectra of graphite. 
Distinction between the C~1s XAS spectra of  Ni@C and HOPG is seen in the insert to \ref{f:C1s-XAS-NiC-multi} 
where the difference spectrum obtained by subtraction of the 
spectrum of HOPG (measured at $\vartheta = 60^{\circ}$) from the spectrum of the as-prepared Ni@C 
sample is presented.  

\begin{figure}[h]
    	\includegraphics[width=0.45\textwidth,angle=270]{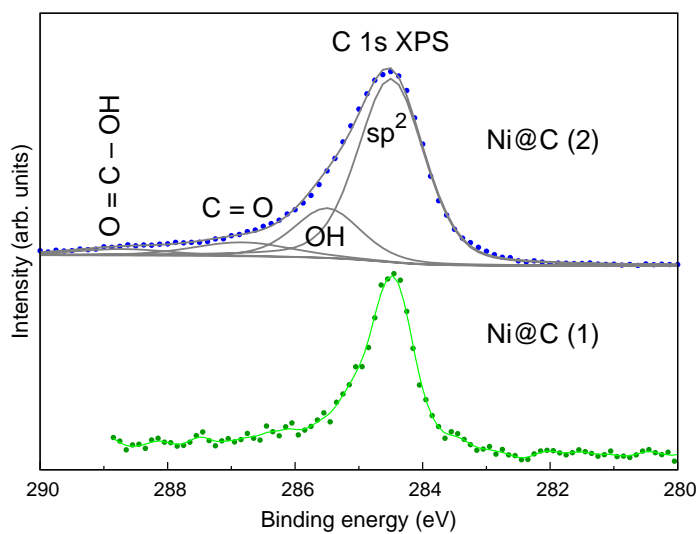}
    	\caption{C~1s X-ray photoelectron spectra of of as-prepared Ni@C nanoparticles (1) 
    	and Ni@C stored in air for 2 years (2).  The spectra of the Ni@C (1) and Ni@C (2) samples are measured 
    	at the excitation energies  of  950~eV and 1486.6~eV, respectively. 
    	}
    	\label{f:XPS-C1s-NiC}
	\end{figure}

The features {\em a} and  {\em b} located below the $\pi^*$ resonance can be seen also in graphene C~1s XAS spectra 
\cite{Entani-2006,Pacile-PRL-2008,Coleman-JPD-2008,Dedkov-10-graphene,Hua-2010}. 
These features are located a little lower than the Fermi level estimated from our C~1s X-ray
photoelectron spectrum (284.5 eV). 
It was suggested \cite{Entani-2006} that the observed transition energy is determined by final state effects. 
Therefore, empty states can be observed below the Fermi level determined by XPS (see \ref{f:XPS-C1s-NiC}). 
An analogous peak in C~1s XAS of  nanographite grains growth on Pt(111) was attributed to a zigzag 
edge-derived electronic state \cite{Entani-2006}. Hua et al. \cite{Hua-2010} showed that this  
peak is determined by defects and should not be present in an ideal graphene sheet. 
Note, an analogous feature at 284.5 eV (peak {\em b})  in the spectra of carbon nanotubes was attributed to 
the rolling (helicity) of the carbon layers of nanotubes \cite{Tang-02}. 

The peaks analogous to  {\em d} and  {\em e} in the spectra of Ni@C can be seen in the spectra of  
graphene \cite{Coleman-JPD-2008} and, according to the work of Hua et al. \cite{Hua-2010}, 
these peaks should be explained by defects.
Both $\pi^*$ and $\sigma^*$ resonances have contributions to these peaks.
Note, that the intense features {\em d} and {\em e} are observed only for the Ni@C sample stored in air for 2 years.
According to our X-ray photoelectron measurements, the sample Ni@C (2) 
is partly oxidized. Therefore, one should assigned these features to contaminations by
 a COOH group  and/or C--H due to  oxidation \cite{Jeong-09-comment}. 

The presence of functional groups in the Ni@C (2) nanoparticles can be confirmed by the C~1s photoelectron 
spectra shown in \ref{f:XPS-C1s-NiC}.
The spectrum of as-prepared Ni@C (1) shows one C~1s peak  at 284.5~eV with an asymmetric line shape 
attributed to ${\rm sp^2}$ bonded carbon atoms  in reference to 
the XPS studies of graphite \cite{perkin-elmer} or carbon nanotubes \cite{Zajickova-2009}. 
The  C~1s XPS spectrum of Ni@C (2) is more complicated  and, in order to separate signals from functional groups, 
we have deconvoluted it into four peaks. As for Ni@C (1), the prominent peak at 284.5~eV reveals C--C interactions. 
The additional peaks in the C~1s XPS spectrum of Ni@C (2) at 285.5~eV, 286.6~eV, and 288.8~eV   can be  
identified as hydroxyl (OH), carbonyl (C$=$O), and carboxyl (O$=$C--OH) groups attached to the carbon atom in Ni@C \cite{Haiber_ABC-2003,Jeong_JACC-2008,Jeong_CPL-08,Jeong_EPL_2008}, respectively, as it is shown
in \ref{f:XPS-C1s-NiC}. The contributions of these functional groups to the C 1s XPS spectrum of Ni@C (2) 
are 19 \%, 8 \%, and 4 \%, respectively. 

\begin{figure}[h]
    	\includegraphics[width=0.40\textwidth,angle=0]{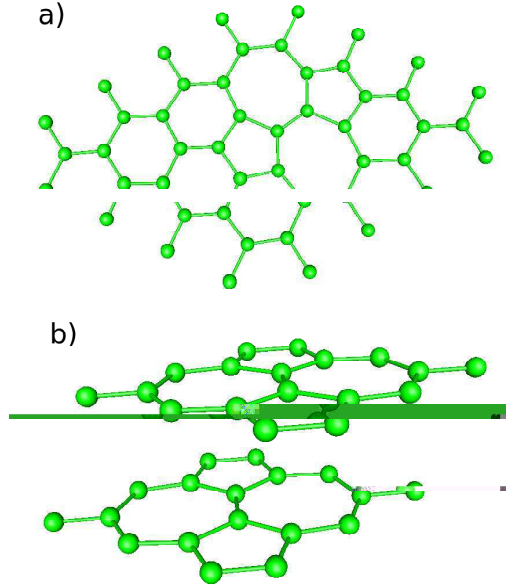}
    	\caption{(a) The geometric structure optimized for graphene monolayer containing one Stone--Wales 
    	defect on supercell with 50 carbon atoms; (b) supercell of graphite with maximal concentration of Stone--Wales defects. }
    	\label{f:structure}
\end{figure}

\begin{figure}[p]
    	\includegraphics[width=0.70\textwidth,angle=0]{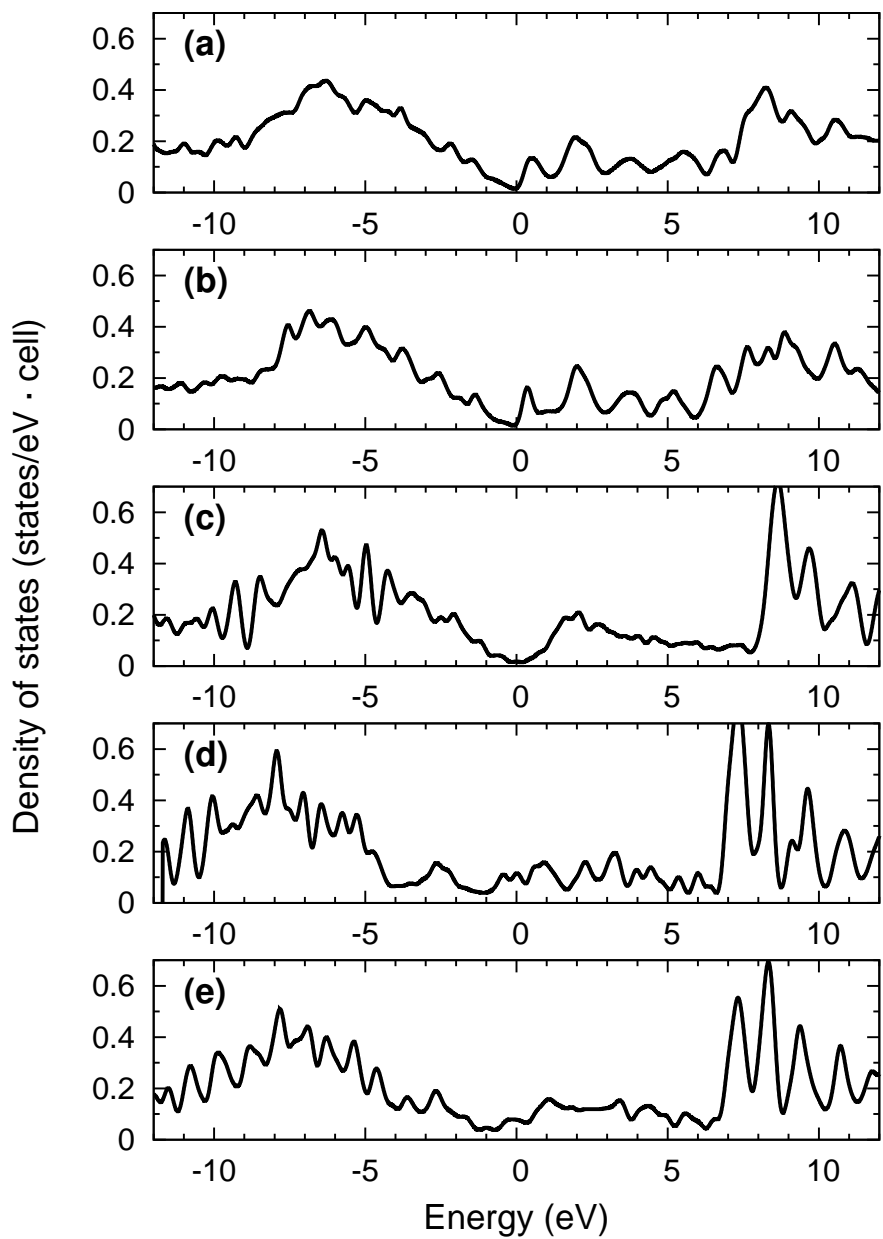}
    	\caption{Densities of states of carbon for graphite and of a few layer of graphene on a single 
    	crystal of Ni(111) with and without Stone--Wales defects: (a)~--- graphite with SW defects; 
    	(b)~--- graphene monolayer with one SW defect;
    	(c)~--- 6 layers of graphene;
    	(d)~--- one layer of graphene on Ni~(111);
    	(e)~--- 3 layers of graphene on Ni (111).
    	}
    	\label{f:DOS}
\end{figure}

Calculations of  the electronic structure of the carbon shell  of Ni@C were carried out for 
a few layer graphene (FLG), i.e., 
6 layers of  graphene with Bernal (AB) stacking,
one and three graphene layers on Ni (111).
The starting geometric structure was taken from the work of Giovannetti et al. \cite{giovannetti-08}.
Stone--Wales (SW) defects \cite{Stone-Wales-86} were used in our model because of  non-planarity of graphene layers 
folded around spherical particles.
In order to examine the role of SW defects for the electronic structure
the calculations for graphene monolayers containing a SW defect 
per 50 atoms in the supercell  [\ref{f:structure} (a)] have been performed.
For imitatation of the discussed defects in graphene multylayers we have used graphite containing 
SW defects in each layer [\ref{f:structure}~(b)]. 
The results of the calculations are presented in \ref{f:DOS}. 

\begin{figure}[h]
    	\includegraphics[width=0.60\textwidth,angle=0]{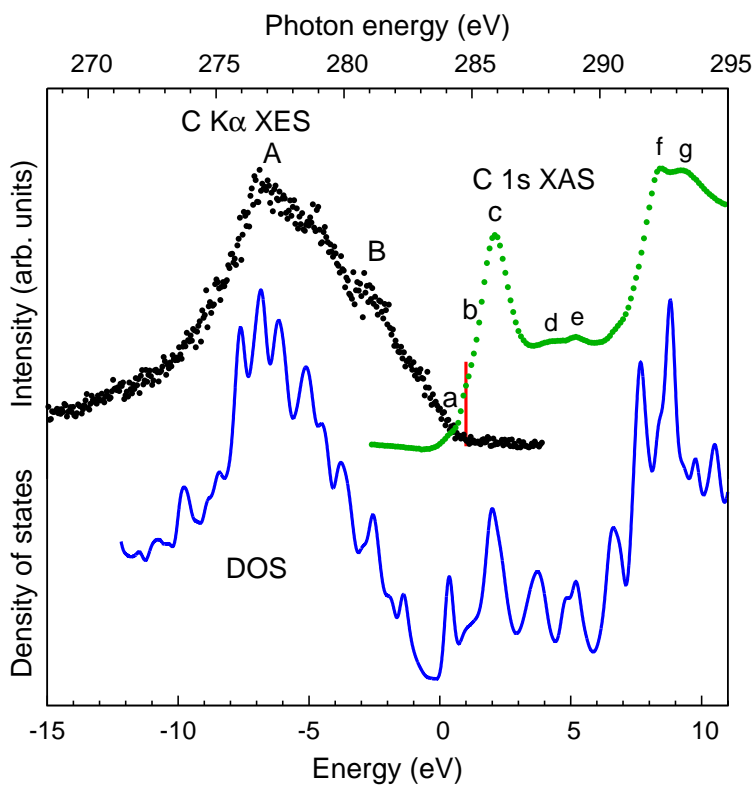}
    	\caption{C K$\alpha$ X-ray emission and C~1s X-ray absorption spectra of as-prepared Ni@C compared with  
    	density of states obtained for 	 layer graphene on a single crystal of Ni(111) with Stone--Wales defects. 
    	The line shows the Fermi level estimated from the C~1s X-ray photoelectron spectrum.}
    	\label{f:C1s-XAS-XES-DOS-NiC}
\end{figure}

For comparison of the theoretical results with our experimental X-ray emission and absorption spectra, we
have simulated the C 2p density of states  of Ni@C as the
sum of DOS for a graphene monolayer with SW defects multiplied by three and 
the DOS for three layers of graphene on Ni(111).   The simulated DOS and experimentally measured
C~K$\alpha$  X-ray emission (XES) and C 1s X-ray absorption (XAS) spectra of Ni@C (1) are presented 
in \ref {f:C1s-XAS-XES-DOS-NiC}.
C~K$\alpha$  X-ray emission spectra reveal  the distribution of C 2p states in the occupied part of the valence band
since they  are appeared as a result of the  $\rm 2p \rightarrow 1s$ electronic transitions.
These XAS and XES spectra for the as-prepared Ni@C sample plotted in the photon energy scale 
are given in the upper part of \ref {f:C1s-XAS-XES-DOS-NiC}. 
The Fermi level estimated from the C~1s X-ray photoelectron spectrum (see \ref{f:XPS-C1s-NiC})
is shown by line.
The  K$\alpha$  X-ray emission spectrum consists of a broad main peak at about 277~eV 
labeled by {\em A}, and a shoulder at about 282 eV labeled by {\em B}. 
Following the interpretation presented in the literature \cite{Skytt-PRB-94,Muramatsu-carbon-2001} one 
can attribute these features to $\sigma$ and $\pi$ orbitals, respectively. 
The combined use of the C~K$\alpha$  X-ray emission and C~1s X-ray absorption spectra (the upper part of 
\ref{f:C1s-XAS-XES-DOS-NiC}) shows that there is no gap between the occupied and unoccupied C 2p states of Ni@C. 

One can see that the calculated superposition of the densities of states  
coincides well with the experimental C K$\alpha$ emission and C 1s absorption spectra.
Two additional features in the C 1s X-ray absorption spectrum near the 
Fermi level (features {\em a} and {\em b} in \ref{f:C1s-XAS-NiC-multi}) 
are caused by the presence of SW defects [feature {\em a} at $+0.5$~eV~--- see also \ref{f:DOS} (a),(b)] and 
overlapping of the electronic states of graphene layers with those of the nickel substrate 
[feature {\em b} at about 1 eV~--- see \ref{f:DOS} (e)].
The comparison of the calculated density of states and C 1s XAS and C K$\alpha$ XES spectra 
shows that the most appropriate structural model for the carbon coating of Ni@C particles 
is graphene multilayers with high concentration of SW defects (one per 18 carbon atoms --- about 6~\%) 
but some electronic contribution from the nearest to the Ni core carbon  shells should be taken into account. 

The similarity of C K$\alpha$ XES and C~1s XAS spectra of Ni@C  (\ref{f:C1s-XAS-XES-DOS-NiC}) 
with those of polymerized fullerene \cite{Boukhvalov-prb-04,Ramm-C60-2000}  
evidences the presence of not only hexagonal but also pentagonal rings of carbon atoms. 
The presence of these defects leads to increased chemical activity of carbon encapsulated metals 
in comparison with that of flat carbon systems \cite{Boukhvalov-nano-08}, 
which is experimentalyl confirmed by  surface oxidation of the Ni@C (2) sample. 
The appearance of Stone--Wales defects caused
by spherical shape of the particles or reconstruction of the defects on the graphene-nickel interface 
was found experimentally \cite{Pablo-09,Camper-graphene-2009} and theoretically \cite{Boukh-destroy-09}. 


\section{Conclusions}\label{sec:conclusion}

In summary, we  have studied the electronic and atomic structures of carbon-coated Ni nanoparticles
by means of X-ray absorption, emission, and photoelectron spectroscopies and DFT calculations.
Results of the measurements suggest that the atomic and electronic structures 
of Ni cores are similar to those of metallic Ni. Absence of any secondary phases and traces of 
oxidation of Ni after two years of exposure of Ni@C nanoparticles in air at ambient 
conditions confirms good protection properties of the carbon shell. 
Measurements of X-ray absorption and photoelectron spectra demostrate some oxidation of the carbon shell. 
Comparison of experimental X-ray spectra and calculated densities of states points to the 
graphite-like structure of the carbon shell with significant amount (about 6\%) of Stone--Wales defects. 
Some overlapping of the electronic states of graphene layers with those of  Ni cores  is also found.


\begin{acknowledgement}

This work is supported by the Russian Foundation for Basic Research  (grants 11-02-00166, 10-02-00323, and 11-02-00022), 
by the Program for Basic Research of the Russian Academy of Sciences 
``Physics of New Materials and Structures'', and by the bilateral  Program ``Russian--German Laboratory at BESSY''.  
DWB acknowledge computational support from the CAC of KIAS. 
\end{acknowledgement}


\begin{mcitethebibliography}{45}
\providecommand*{\natexlab}[1]{#1}
\providecommand*{\mciteSetBstSublistMode}[1]{}
\providecommand*{\mciteSetBstMaxWidthForm}[2]{}
\providecommand*{\mciteBstWouldAddEndPuncttrue}
  {\def\EndOfBibitem{\unskip.}}
\providecommand*{\mciteBstWouldAddEndPunctfalse}
  {\let\EndOfBibitem\relax}
\providecommand*{\mciteSetBstMidEndSepPunct}[3]{}
\providecommand*{\mciteSetBstSublistLabelBeginEnd}[3]{}
\providecommand*{\EndOfBibitem}{}
\mciteSetBstSublistMode{f}
\mciteSetBstMaxWidthForm{subitem}{(\alph{mcitesubitemcount})}
\mciteSetBstSublistLabelBeginEnd{\mcitemaxwidthsubitemform\space}
{\relax}{\relax}

\bibitem[Wu et~al.(2002)Wu, Liu, Liu, Haley, Treadway, Larson, Ge, Peale, and
  Bruchez]{Wu-Immunofluorescent-02}
Wu,~X.; Liu,~H.; Liu,~J.; Haley,~K.~N.; Treadway,~J.~A.; Larson,~J.~P.; Ge,~N.;
  Peale,~F.; Bruchez,~M.~P. \emph{Nature Biotechnology} \textbf{2002},
  \emph{21}, 41--46\relax
\mciteBstWouldAddEndPuncttrue
\mciteSetBstMidEndSepPunct{\mcitedefaultmidpunct}
{\mcitedefaultendpunct}{\mcitedefaultseppunct}\relax
\EndOfBibitem
\bibitem[Paciotti et~al.(2004)Paciotti, Myer, Weinreich, Goia, Pavel,
  McLaughlin, and Tamarkin]{Paciott-04-nanoparticle}
Paciotti,~G.; Myer,~L.; Weinreich,~D.; Goia,~D.; Pavel,~N.; McLaughlin,~R.~E.;
  Tamarkin,~L. \emph{Drug Deliv.} \textbf{2004}, \emph{11}, 169--183\relax
\mciteBstWouldAddEndPuncttrue
\mciteSetBstMidEndSepPunct{\mcitedefaultmidpunct}
{\mcitedefaultendpunct}{\mcitedefaultseppunct}\relax
\EndOfBibitem
\bibitem[Alivisatos(2004)]{Alivisatos-04-nano-bio}
Alivisatos,~P. \emph{Nature Biotechnology} \textbf{2004}, \emph{22},
  47--52\relax
\mciteBstWouldAddEndPuncttrue
\mciteSetBstMidEndSepPunct{\mcitedefaultmidpunct}
{\mcitedefaultendpunct}{\mcitedefaultseppunct}\relax
\EndOfBibitem
\bibitem[Seo et~al.(2006)Seo, Lee, Sun, Suzuki, Mann, Liu, Terashima, Yang,
  Mcconnell, Nishimura, and Dai]{Seo_FeCo-C-nano-2006}
Seo,~W.~S.; Lee,~J.~H.; Sun,~X.; Suzuki,~Y.; Mann,~D.; Liu,~Z.; Terashima,~M.;
  Yang,~P.~C.; Mcconnell,~M.~V.; Nishimura,~D.~G.; Dai,~H. \emph{Nature Mat.}
  \textbf{2006}, \emph{5}, 971--976\relax
\mciteBstWouldAddEndPuncttrue
\mciteSetBstMidEndSepPunct{\mcitedefaultmidpunct}
{\mcitedefaultendpunct}{\mcitedefaultseppunct}\relax
\EndOfBibitem
\bibitem[Hayashi et~al.(2006)Hayashi, Tokunaga, Kaneko, S.~J.~Henley, Carey,
  and Silva]{Hayashi-06-carbon-nanotubes}
Hayashi,~Y.; Tokunaga,~T.; Kaneko,~K.; S.~J.~Henley,~V.~S.; Carey,~J.~D.;
  Silva,~S.~R.~P. \emph{IEEE Trans. on Nanotechnology} \textbf{2006}, \emph{6},
  485--490\relax
\mciteBstWouldAddEndPuncttrue
\mciteSetBstMidEndSepPunct{\mcitedefaultmidpunct}
{\mcitedefaultendpunct}{\mcitedefaultseppunct}\relax
\EndOfBibitem
\bibitem[El\'{i}as et~al.(2005)El\'{i}as, Rodr\'{i}guez-Manzo, McCartney,
  Golberg, Zamudio, Baltazar, L\'{o}pez-Ur\'{i}as, {Mu\~{n}oz-Sandoval}, Gu,
  Tang, Smith, Bando, Terrones, and Terrones]{Elias-05-FeCo-nanowires}
El\'{i}as,~A.~L.; Rodr\'{i}guez-Manzo,~J.~A.; McCartney,~M.~R.; Golberg,~D.;
  Zamudio,~A.; Baltazar,~S.~E.; L\'{o}pez-Ur\'{i}as,~F.;
  {Mu\~{n}oz-Sandoval},~E.; Gu,~L.; Tang,~C.~C.; Smith,~D.~J.; Bando,~Y.;
  Terrones,~H.; Terrones,~M. \emph{Nano Lett.} \textbf{2005}, \emph{5},
  467--472\relax
\mciteBstWouldAddEndPuncttrue
\mciteSetBstMidEndSepPunct{\mcitedefaultmidpunct}
{\mcitedefaultendpunct}{\mcitedefaultseppunct}\relax
\EndOfBibitem
\bibitem[Leonhardt et~al.(2003)Leonhardt, Ritschel, Kozhuharova, Graffa,
  M\"{u}hl, Huhle, M\"{o}nch, Elefant, and Schneider]{Leonhardt-03-nanotubes}
Leonhardt,~A.; Ritschel,~M.; Kozhuharova,~R.; Graffa,~A.; M\"{u}hl,~T.;
  Huhle,~R.; M\"{o}nch,~I.; Elefant,~D.; Schneider,~C.~M. \emph{Diamond Relat.
  Mater.} \textbf{2003}, \emph{12}, 790--793\relax
\mciteBstWouldAddEndPuncttrue
\mciteSetBstMidEndSepPunct{\mcitedefaultmidpunct}
{\mcitedefaultendpunct}{\mcitedefaultseppunct}\relax
\EndOfBibitem
\bibitem[Choi et~al.(2003)Choi, Kang, and Hwang]{Choi-03-Cu-nanowires}
Choi,~W.~Y.; Kang,~J.~W.; Hwang,~H.~J. \emph{Phys. Rev. B} \textbf{2003},
  \emph{68}, 193405\relax
\mciteBstWouldAddEndPuncttrue
\mciteSetBstMidEndSepPunct{\mcitedefaultmidpunct}
{\mcitedefaultendpunct}{\mcitedefaultseppunct}\relax
\EndOfBibitem
\bibitem[Bao et~al.(2002)Bao, Tie, Xu, Suo, Zhou, and Hong]{Bao-02-nanotubes}
Bao,~J.; Tie,~C.; Xu,~Z.; Suo,~Z.; Zhou,~Q.; Hong,~J. \emph{Adv. Mater.}
  \textbf{2002}, \emph{14}, 1483--1486\relax
\mciteBstWouldAddEndPuncttrue
\mciteSetBstMidEndSepPunct{\mcitedefaultmidpunct}
{\mcitedefaultendpunct}{\mcitedefaultseppunct}\relax
\EndOfBibitem
\bibitem[Bao et~al.(2002)Bao, Zhou, Hong, and Xu]{Bao-02-Ni-nano-magn}
Bao,~J.; Zhou,~Q.; Hong,~J.; Xu,~Z. \emph{Appl. Phys. Lett.} \textbf{2002},
  \emph{81}, 4592--4594\relax
\mciteBstWouldAddEndPuncttrue
\mciteSetBstMidEndSepPunct{\mcitedefaultmidpunct}
{\mcitedefaultendpunct}{\mcitedefaultseppunct}\relax
\EndOfBibitem
\bibitem[Scott and Majetich(1995)]{Scott-95-nano-morphology}
Scott,~J. H.~J.; Majetich,~S.~A. \emph{Phys. Rev. B} \textbf{1995}, \emph{52},
  12564--12571\relax
\mciteBstWouldAddEndPuncttrue
\mciteSetBstMidEndSepPunct{\mcitedefaultmidpunct}
{\mcitedefaultendpunct}{\mcitedefaultseppunct}\relax
\EndOfBibitem
\bibitem[Ruoff et~al.(1993)Ruoff, Lorents, Chan, Malhotra, and
  Subramoney]{Ruoff_LaC-nature-1993}
Ruoff,~R.~S.; Lorents,~D.~C.; Chan,~B.; Malhotra,~R.; Subramoney,~S.
  \emph{Science} \textbf{1993}, \emph{259}, 346--348\relax
\mciteBstWouldAddEndPuncttrue
\mciteSetBstMidEndSepPunct{\mcitedefaultmidpunct}
{\mcitedefaultendpunct}{\mcitedefaultseppunct}\relax
\EndOfBibitem
\bibitem[Lokteva et~al.(2009)Lokteva, Kachevskii, Turakulova, Golubina, Lunin,
  Ermakov, Uimin, and Mysik]{lokteva-2009}
Lokteva,~E.; Kachevskii,~S.~A.; Turakulova,~A.~O.; Golubina,~E.~V.;
  Lunin,~V.~V.; Ermakov,~A.~E.; Uimin,~M.~A.; Mysik,~A.~A. \emph{Russian J.
  Phys. Chem. A} \textbf{2009}, \emph{83}, 1300--1306\relax
\mciteBstWouldAddEndPuncttrue
\mciteSetBstMidEndSepPunct{\mcitedefaultmidpunct}
{\mcitedefaultendpunct}{\mcitedefaultseppunct}\relax
\EndOfBibitem
\bibitem[Ermakov et~al.(2009)Ermakov, Uimin, Lokteva, Mysik, Kachevskii,
  Turakulova, Gaviko, and Lunin]{ermakov-RJPC-2009}
Ermakov,~A.~E.; Uimin,~M.~A.; Lokteva,~E.~S.; Mysik,~A.~A.; Kachevskii,~S.~A.;
  Turakulova,~A.~O.; Gaviko,~V.~S.; Lunin,~V.~V. \emph{Russian J. Phys. Chem.
  A} \textbf{2009}, \emph{83}, 1187--1193\relax
\mciteBstWouldAddEndPuncttrue
\mciteSetBstMidEndSepPunct{\mcitedefaultmidpunct}
{\mcitedefaultendpunct}{\mcitedefaultseppunct}\relax
\EndOfBibitem
\bibitem[Huang et~al.(2009)Huang, Xu, Yang, Tang, Huang, , and
  Shen]{Huang-NiC-2009}
Huang,~Y.; Xu,~Z.; Yang,~Y.; Tang,~T.; Huang,~R.; ; Shen,~J. \emph{J. Phys.
  Chem. C} \textbf{2009}, \emph{113}, 6533--6538\relax
\mciteBstWouldAddEndPuncttrue
\mciteSetBstMidEndSepPunct{\mcitedefaultmidpunct}
{\mcitedefaultendpunct}{\mcitedefaultseppunct}\relax
\EndOfBibitem
\bibitem[Galakhov et~al.(2010)Galakhov, Shkvarin, Semenova, Uimin, Mysik,
  Shchegoleva, Yermakov, and Kurmaev]{Galakhov-JPCC-2010}
Galakhov,~V.~R.; Shkvarin,~A.~S.; Semenova,~A.~S.; Uimin,~M.~A.; Mysik,~A.~A.;
  Shchegoleva,~N.~N.; Yermakov,~A.~Y.; Kurmaev,~E.~Z. \emph{J. Phys. Chem. C}
  \textbf{2010}, \emph{114}, 22413--22416\relax
\mciteBstWouldAddEndPuncttrue
\mciteSetBstMidEndSepPunct{\mcitedefaultmidpunct}
{\mcitedefaultendpunct}{\mcitedefaultseppunct}\relax
\EndOfBibitem
\bibitem[Artacho et~al.(2004)Artacho, Gale, Garc\'{i}a, Junquera, Martin,
  Ordej\'{o}n, S\'{a}nchez-Portal, and Soler]{SIESTA}
Artacho,~E.; Gale,~J.~D.; Garc\'{i}a,~A.; Junquera,~J.; Martin,~R.~M.;
  Ordej\'{o}n,~P.; S\'{a}nchez-Portal,~D.; Soler,~J.~M. \emph{SIESTA, Version
  1.3};
\newblock 2004\relax
\mciteBstWouldAddEndPuncttrue
\mciteSetBstMidEndSepPunct{\mcitedefaultmidpunct}
{\mcitedefaultendpunct}{\mcitedefaultseppunct}\relax
\EndOfBibitem
\bibitem[Soler et~al.(2002)Soler, Artacho, Gale, Garc\'{i}a, Junquera,
  Ordej\'{o}n, and S\'{a}nchez-Portal]{Soler-31}
Soler,~J.~M.; Artacho,~E.; Gale,~J.~D.; Garc\'{i}a,~A.; Junquera,~J.;
  Ordej\'{o}n,~P.; S\'{a}nchez-Portal,~D. \emph{J. Phys.: Condens. Matter}
  \textbf{2002}, \emph{14}, 2475\relax
\mciteBstWouldAddEndPuncttrue
\mciteSetBstMidEndSepPunct{\mcitedefaultmidpunct}
{\mcitedefaultendpunct}{\mcitedefaultseppunct}\relax
\EndOfBibitem
\bibitem[Perdew and Zunger(1981)]{perdew_1981}
Perdew,~J.~P.; Zunger,~A. \emph{Phys.\ Rev.\ B} \textbf{1981}, \emph{77},
  5048--5079\relax
\mciteBstWouldAddEndPuncttrue
\mciteSetBstMidEndSepPunct{\mcitedefaultmidpunct}
{\mcitedefaultendpunct}{\mcitedefaultseppunct}\relax
\EndOfBibitem
\bibitem[Yeh and Lindau(1985)]{yeh85}
Yeh,~J.~J.; Lindau,~I. \emph{At. Data Nucl. Data Tables} \textbf{1985},
  \emph{32}, 1--155\relax
\mciteBstWouldAddEndPuncttrue
\mciteSetBstMidEndSepPunct{\mcitedefaultmidpunct}
{\mcitedefaultendpunct}{\mcitedefaultseppunct}\relax
\EndOfBibitem
\bibitem[Cho et~al.(2004)Cho, Kauzlarich, Olamit, Liu, Grandjean, Rebbouh, and
  Long]{Cho_Fe-Au-2004}
Cho,~S.-J.; Kauzlarich,~S.~M.; Olamit,~J.; Liu,~K.; Grandjean,~F.; Rebbouh,~L.;
  Long,~G.~J. \emph{J. Appl. Phys.} \textbf{2004}, \emph{95}, 6804--6805\relax
\mciteBstWouldAddEndPuncttrue
\mciteSetBstMidEndSepPunct{\mcitedefaultmidpunct}
{\mcitedefaultendpunct}{\mcitedefaultseppunct}\relax
\EndOfBibitem
\bibitem[Entani et~al.(2006)Entani, Ikeda, Kiguchi, and Saiki]{Entani-2006}
Entani,~S.; Ikeda,~S.; Kiguchi,~M.; Saiki,~K. \emph{Appl. Phys. Lett.}
  \textbf{2006}, \emph{88}, 153126\relax
\mciteBstWouldAddEndPuncttrue
\mciteSetBstMidEndSepPunct{\mcitedefaultmidpunct}
{\mcitedefaultendpunct}{\mcitedefaultseppunct}\relax
\EndOfBibitem
\bibitem[Pacil{\'e} et~al.(2008)Pacil{\'e}, Papagno, Rodr\'{i}guez, Grioni,
  Papagno, Girit, Meyer, Begtrup, and Zettl]{Pacile-PRL-2008}
Pacil{\'e},~D.; Papagno,~M.; Rodr\'{i}guez,~A.~F.; Grioni,~M.; Papagno,~L.;
  Girit,~{\k{C}}.~{\" {O}}.; Meyer,~J.~C.; Begtrup,~G.~E.; Zettl,~A.
  \emph{Phys. Rev. Lett.} \textbf{2008}, \emph{101}, 066806\relax
\mciteBstWouldAddEndPuncttrue
\mciteSetBstMidEndSepPunct{\mcitedefaultmidpunct}
{\mcitedefaultendpunct}{\mcitedefaultseppunct}\relax
\EndOfBibitem
\bibitem[Coleman et~al.(2008)Coleman, Knut, Karis, Grennberg, Jansson, Quinlan,
  Holloway, Sanyal, and Eriksson]{Coleman-JPD-2008}
Coleman,~V.~A.; Knut,~R.; Karis,~O.; Grennberg,~H.; Jansson,~U.; Quinlan,~R.;
  Holloway,~B.~C.; Sanyal,~B.; Eriksson,~O. \emph{J. Phys. D: Appl. Phys.}
  \textbf{2008}, \emph{41}, 062001\relax
\mciteBstWouldAddEndPuncttrue
\mciteSetBstMidEndSepPunct{\mcitedefaultmidpunct}
{\mcitedefaultendpunct}{\mcitedefaultseppunct}\relax
\EndOfBibitem
\bibitem[Dedkov et~al.(2010)Dedkov, Sicot, and Fonin]{Dedkov-10-graphene}
Dedkov,~Y.~S.; Sicot,~M.; Fonin,~M. \emph{J. Appl. Phys.} \textbf{2010},
  \emph{107}, 09E121\relax
\mciteBstWouldAddEndPuncttrue
\mciteSetBstMidEndSepPunct{\mcitedefaultmidpunct}
{\mcitedefaultendpunct}{\mcitedefaultseppunct}\relax
\EndOfBibitem
\bibitem[Hua et~al.(2010)Hua, Gao, Li, {\AA}gren, and Luo]{Hua-2010}
Hua,~W.; Gao,~B.; Li,~S.; {\AA}gren,~H.; Luo,~Y. \emph{Phys. Rev. B}
  \textbf{2010}, \emph{82}, 155433\relax
\mciteBstWouldAddEndPuncttrue
\mciteSetBstMidEndSepPunct{\mcitedefaultmidpunct}
{\mcitedefaultendpunct}{\mcitedefaultseppunct}\relax
\EndOfBibitem
\bibitem[Tang et~al.(2002)Tang, Sham, Hu, Lee, and Lee]{Tang-02}
Tang,~Y.~H.; Sham,~T.~K.; Hu,~Y.~F.; Lee,~C.~S.; Lee,~S.~T. \emph{Chem. Phys.
  Lett.} \textbf{2002}, \emph{366}, 636--641\relax
\mciteBstWouldAddEndPuncttrue
\mciteSetBstMidEndSepPunct{\mcitedefaultmidpunct}
{\mcitedefaultendpunct}{\mcitedefaultseppunct}\relax
\EndOfBibitem
\bibitem[Jeong et~al.(2009)Jeong, Noh, Kim, Glans, Jin, Smith, and
  Lee]{Jeong-09-comment}
Jeong,~H.-K.; Noh,~H.-J.; Kim,~J.-Y.; Glans,~P.-A.; Jin,~M.~H.; Smith,~K.~E.;
  Lee,~Y.~H. \emph{Phys. Rev. Lett.} \textbf{2009}, \emph{102}, 099701\relax
\mciteBstWouldAddEndPuncttrue
\mciteSetBstMidEndSepPunct{\mcitedefaultmidpunct}
{\mcitedefaultendpunct}{\mcitedefaultseppunct}\relax
\EndOfBibitem
\bibitem[Chastain(1992)]{perkin-elmer}
Chastain,~J. \emph{Handbook of X-ray Photoelectron Spectroscopy};
\newblock Eden Prairie: Perkin Elmer Corporation, 1992\relax
\mciteBstWouldAddEndPuncttrue
\mciteSetBstMidEndSepPunct{\mcitedefaultmidpunct}
{\mcitedefaultendpunct}{\mcitedefaultseppunct}\relax
\EndOfBibitem
\bibitem[Zaj\'{i}\v{c}kova et~al.(2009)Zaj\'{i}\v{c}kova, Ku\v{c}erov\'{a},
  Bur\v{s}ikov\'{a}, Eli\'{a}\v{s}, Houdkov\'{a}, Synek, Mar\v{s}\'{i}kov\'{a},
  and Ja\v{s}ek]{Zajickova-2009}
Zaj\'{i}\v{c}kova,~L.; Ku\v{c}erov\'{a},~Z.; Bur\v{s}ikov\'{a},~V.;
  Eli\'{a}\v{s},~M.; Houdkov\'{a},~J.; Synek,~P.; Mar\v{s}\'{i}kov\'{a},~H.;
  Ja\v{s}ek,~O. \emph{Plasma Process. Polymers} \textbf{2009}, \emph{6},
  S864--S869\relax
\mciteBstWouldAddEndPuncttrue
\mciteSetBstMidEndSepPunct{\mcitedefaultmidpunct}
{\mcitedefaultendpunct}{\mcitedefaultseppunct}\relax
\EndOfBibitem
\bibitem[Haiber et~al.(2003)Haiber, Ai, Bubert, Heintze, Br{\"{u}}ser, Brandl,
  and Marginean]{Haiber_ABC-2003}
Haiber,~S.; Ai,~X.; Bubert,~H.; Heintze,~M.; Br{\"{u}}ser,~V.; Brandl,~W.;
  Marginean,~G. \emph{Anal. Bioanal. Chem.} \textbf{2003}, \emph{130},
  875--883\relax
\mciteBstWouldAddEndPuncttrue
\mciteSetBstMidEndSepPunct{\mcitedefaultmidpunct}
{\mcitedefaultendpunct}{\mcitedefaultseppunct}\relax
\EndOfBibitem
\bibitem[Jeong et~al.(2008)Jeong, Lee, Lahaye, Park, An, Kim, Yang, Park,
  Ruoff, and Lee]{Jeong_JACC-2008}
Jeong,~H.-K.; Lee,~Y.~P.; Lahaye,~R. J. W.~E.; Park,~M.-H.; An,~K.~H.;
  Kim,~I.~J.; Yang,~C.-W.; Park,~C.~Y.; Ruoff,~R.~S.; Lee,~Y.~H. \emph{J. Am.
  Chem. Soc.} \textbf{2008}, \emph{130}, 1362--1366\relax
\mciteBstWouldAddEndPuncttrue
\mciteSetBstMidEndSepPunct{\mcitedefaultmidpunct}
{\mcitedefaultendpunct}{\mcitedefaultseppunct}\relax
\EndOfBibitem
\bibitem[Jeong et~al.(2008)Jeong, Colakerol, Jin, Glans, Smith, and
  Lee]{Jeong_CPL-08}
Jeong,~H.-K.; Colakerol,~L.; Jin,~M.~H.; Glans,~P.-A.; Smith,~K.~E.; Lee,~Y.~H.
  \emph{Chem. Phys. Lett.} \textbf{2008}, \emph{460}, 499--502\relax
\mciteBstWouldAddEndPuncttrue
\mciteSetBstMidEndSepPunct{\mcitedefaultmidpunct}
{\mcitedefaultendpunct}{\mcitedefaultseppunct}\relax
\EndOfBibitem
\bibitem[Jeong et~al.(2008)Jeong, Noh, Kim, Jin, Park, and Lee]{Jeong_EPL_2008}
Jeong,~H.-K.; Noh,~H.-J.; Kim,~J.-Y.; Jin,~M.~H.; Park,~C.~Y.; Lee,~Y.~H.
  \emph{Europhys. Lett.} \textbf{2008}, \emph{82}, 67004\relax
\mciteBstWouldAddEndPuncttrue
\mciteSetBstMidEndSepPunct{\mcitedefaultmidpunct}
{\mcitedefaultendpunct}{\mcitedefaultseppunct}\relax
\EndOfBibitem
\bibitem[Giovannetti et~al.(2008)Giovannetti, Khomyakov, Brocks, Karpan,
  van~den Brinkand, and Kelly]{giovannetti-08}
Giovannetti,~G.; Khomyakov,~P.~A.; Brocks,~G.; Karpan,~V.~M.; van~den
  Brinkand,~J.; Kelly,~P.~J. \emph{Phys. Rev. Lett.} \textbf{2008}, \emph{101},
  026803\relax
\mciteBstWouldAddEndPuncttrue
\mciteSetBstMidEndSepPunct{\mcitedefaultmidpunct}
{\mcitedefaultendpunct}{\mcitedefaultseppunct}\relax
\EndOfBibitem
\bibitem[Stone and Wales(1986)]{Stone-Wales-86}
Stone,~A.~J.; Wales,~D.~J. \emph{Chem. Phys. Lett.} \textbf{1986}, \emph{128},
  501--503\relax
\mciteBstWouldAddEndPuncttrue
\mciteSetBstMidEndSepPunct{\mcitedefaultmidpunct}
{\mcitedefaultendpunct}{\mcitedefaultseppunct}\relax
\EndOfBibitem
\bibitem[Skytt et~al.(1994)Skytt, Glans, Mancini, Guo, Wassdahl, Nordgren, and
  Ma]{Skytt-PRB-94}
Skytt,~P.; Glans,~P.; Mancini,~D.~C.; Guo,~J.-H.; Wassdahl,~N.; Nordgren,~J.;
  Ma,~Y. \emph{Phys. Rev. B} \textbf{1994}, \emph{50}, 10457--10461\relax
\mciteBstWouldAddEndPuncttrue
\mciteSetBstMidEndSepPunct{\mcitedefaultmidpunct}
{\mcitedefaultendpunct}{\mcitedefaultseppunct}\relax
\EndOfBibitem
\bibitem[Muramatsu et~al.(2001)Muramatsu, Hirono, Umemura, Ueno, Hayashi,
  Grush, Gullikson, and Perera]{Muramatsu-carbon-2001}
Muramatsu,~Y.; Hirono,~S.; Umemura,~S.; Ueno,~Y.; Hayashi,~T.; Grush,~M.~M.;
  Gullikson,~E.~M.; Perera,~R. C.~C. \emph{Carbon} \textbf{2001}, \emph{39},
  1403--1407\relax
\mciteBstWouldAddEndPuncttrue
\mciteSetBstMidEndSepPunct{\mcitedefaultmidpunct}
{\mcitedefaultendpunct}{\mcitedefaultseppunct}\relax
\EndOfBibitem
\bibitem[Boukhvalov et~al.(2004)Boukhvalov, Karimov, Kurmaev, Hamilton, Moewes,
  Finkelstein, Katsnelson, Davydov, Rakhmanina, Makarova, Kopelevich,
  Chiuzbaian, and Neumann]{Boukhvalov-prb-04}
Boukhvalov,~D.~W.; Karimov,~P.~F.; Kurmaev,~E.~Z.; Hamilton,~T.; Moewes,~A.;
  Finkelstein,~L.~D.; Katsnelson,~M.~I.; Davydov,~V.~A.; Rakhmanina,~A.~V.;
  Makarova,~T.~L.; Kopelevich,~Y.; Chiuzbaian,~S.; Neumann,~M. \emph{Phys. Rev.
  B} \textbf{2004}, \emph{69}, 115425\relax
\mciteBstWouldAddEndPuncttrue
\mciteSetBstMidEndSepPunct{\mcitedefaultmidpunct}
{\mcitedefaultendpunct}{\mcitedefaultseppunct}\relax
\EndOfBibitem
\bibitem[Ramm et~al.(2000)Ramm, Ata, Gross, and Unger]{Ramm-C60-2000}
Ramm,~M.; Ata,~M.; Gross,~T.; Unger,~W. \emph{Appl. Phys. A} \textbf{2000},
  \emph{70}, 387--390\relax
\mciteBstWouldAddEndPuncttrue
\mciteSetBstMidEndSepPunct{\mcitedefaultmidpunct}
{\mcitedefaultendpunct}{\mcitedefaultseppunct}\relax
\EndOfBibitem
\bibitem[Boukhvalov and Katsnelson(2008)]{Boukhvalov-nano-08}
Boukhvalov,~D.~W.; Katsnelson,~M.~I. \emph{Nano Lett.} \textbf{2008}, \emph{8},
  4373--4379\relax
\mciteBstWouldAddEndPuncttrue
\mciteSetBstMidEndSepPunct{\mcitedefaultmidpunct}
{\mcitedefaultendpunct}{\mcitedefaultseppunct}\relax
\EndOfBibitem
\bibitem[Lemme et~al.(2009)Lemme, Bell, Williams, Stern, Baugher,
  Jarillo-Herrero, and Marcus]{Pablo-09}
Lemme,~M.~C.; Bell,~D.~C.; Williams,~J.~R.; Stern,~L.~A.; Baugher,~B. W.~H.;
  Jarillo-Herrero,~P.; Marcus,~C.~M. \emph{ACS Nano} \textbf{2009}, \emph{3},
  2674--2676\relax
\mciteBstWouldAddEndPuncttrue
\mciteSetBstMidEndSepPunct{\mcitedefaultmidpunct}
{\mcitedefaultendpunct}{\mcitedefaultseppunct}\relax
\EndOfBibitem
\bibitem[Campos et~al.(2009)Campos, Manfrinato, Sanchez-Yamagishi, Kong, and
  Jarillo-Herrero]{Camper-graphene-2009}
Campos,~L.~C.; Manfrinato,~V.~R.; Sanchez-Yamagishi,~J.~D.; Kong,~J.;
  Jarillo-Herrero,~P. \emph{Nano Lett.} \textbf{2009}, \emph{9},
  2600--2604\relax
\mciteBstWouldAddEndPuncttrue
\mciteSetBstMidEndSepPunct{\mcitedefaultmidpunct}
{\mcitedefaultendpunct}{\mcitedefaultseppunct}\relax
\EndOfBibitem
\bibitem[Boukhvalov and Katsnelson(2009)]{Boukh-destroy-09}
Boukhvalov,~D.~W.; Katsnelson,~M.~I. \emph{Appl. Phys. Lett.} \textbf{2009},
  \emph{95}, 023109\relax
\mciteBstWouldAddEndPuncttrue
\mciteSetBstMidEndSepPunct{\mcitedefaultmidpunct}
{\mcitedefaultendpunct}{\mcitedefaultseppunct}\relax
\EndOfBibitem
\end{mcitethebibliography}
\providecommand*{\mcitethebibliography}{\thebibliography}
\csname @ifundefined\endcsname{endmcitethebibliography}
{\let\endmcitethebibliography\endthebibliography}{}

\end{document}